# Determination of weak squeezed vacuum state through photon statistics measurement

Guanhua Zuo [a], Yuchi Zhang [b], Jing Li [b], Shiyao Zhu [c], Yanqiang Guo [d, †], Tiancai Zhang [a, ‡]

a *State Key Laboratory of Quantum Qptics and Quantum Optics Devices, Collaborative Innovation Center of Extreme Optics, Institute of Opto-Electronics, Shanxi University, Taiyuan 030006, China*

b *College of Physics and Electronic Engineering, Shanxi University, Taiyuan 030006, China*

c *Department of Physics, Zhejiang University, Hangzhou 310027, China*

d *Key Laboratory of Advanced Transducers and Intelligent Control System, Ministry of Education and Shanxi Province, College of Physics and Optoelectronics, Taiyuan University of Technology, Taiyuan 030600, China*

†Corresponding author at: Key Laboratory of Advanced Transducers and Intelligent Control System, Ministry of Education and Shanxi Province, College of Physics and Optoelectronics, Taiyuan University of Technology, Taiyuan 030600, China

‡ Corresponding author at: State Key Laboratory of Quantum Qptics and Quantum Optics Devices, Collaborative Innovation Center of Extreme Optics, Institute of Opto-Electronics, Shanxi University, Taiyuan 030006, China

E-mail addresses: guoyanqiang@tyut.edu.cn (Yanqiang Guo), tczhang@sxu.edu.cn (Tiancai Zhang).

## Abstract

Weak squeezed vacuum light, especially resonant to the atomic transition, plays an important role in quantum storage and generation of various quantum sources. However, the general homodyne detection (HD) cannot determine weak squeezing due to the low signal to noise ratio and the limited resolution of the HD system. Here we provide an alternative method based on photon statistics measurement to determine the weak squeezing of the squeezed vacuum light generated from an optical parametric oscillator working far below the threshold. The approach is established the relationship between the squeezing parameter and the second-order degree of coherence. The theoretical analysis agrees well with the experiment results. The advantage of this method is that it provides a feasible and reliable experimental measure to determine the weak squeezing with high precision and the measurement is independent on the detection efficiency. This method can be used to measure other quantum features for various quantum states with extremely weak non-classicality.

1. **Introduction**

As a well investigated nonclassical light, optical squeezed state is an important source for many applications in quantum information[1-4], precision measurement[5-7], quantum communication[8-10], and the interaction between light and atoms[11-13]. Up to now, people have already done lots of studies on the generation, detection and application of the squeezed light, and various squeezed states have been generated in experiments, such as polarization squeezed state[14-15], photon number squeezed state[16], and higher-order squeezed state[17]. For different applications, people have generated various wavebands from communication wavelength to those wavelengths corresponding to specific atomic transition[18-19] and different analysis frequency bands, such as frequency-dependent squeezing for gravitational wave detection[20], low-frequency squeezing for precision measurement of magnetic field[21], and high-frequency squeezing for time-domain multiplexing in quantum communication[22]. Currently, the highest squeezing of 15 dB is obtained and it has significant applications in quantum metrology[23].

Contrary to the high squeezing obtained near the threshold of an optical parametric oscillator (OPO), a weak squeezing can be generated when the OPO operates far below the threshold. The weak squeezed vacuum light shows strong photon bunching effect[24-25]and that feature can be used for enhancing multiphoton nonlinear light-matter interaction[26] and improving visibility of ghost interference and ghost imaging[27-28]. Meanwhile, the weak squeezed vacuum light, especially resonant to the atomic transition, plays a key role in generation of various quantum sources[29], such as single-photon state[30], photon-pair state[31], and Schrodinger kittens state[32]. The weak squeezing can also contribute to the applications for quantum communication[33]. Moreover, weak displaced-squeezed states exhibiting photon antibunching effect has been prepared and measured, which characterizes a nonclassical statistics of the weak squeezed state[34]. In the above mentioned experiments, the squeezing parameter is a key factor to affect the performance of the quantum state preparation and the non-classicality which are fundamental to quantum communication and precision measurement. It is thus very important to determine the squeezing parameter (i.e., squeezing degree) in the weak pump regime. We know that the squeezing is normally measured by using a usual homodyne detection (HD) [35]. However, it is very difficult or even impossible to determine the weak squeezing due to the low signal to noise ratio and limited resolution of the HD system. On the other hand, the measurement of HD is quite affected by the total detection efficiency and low efficiency even deteriorates the squeezing. The determination of the weak squeezing still remains to be explored.

In this paper, we have studied the determination of weak squeezing in experiment and theory. We theoretically established the quantitative relationship between the weak squeezing and the second order degree of coherence in the OPO process. It shows that if the second-order degree of coherence and the parameters of the OPO cavity itself can be measured, the weak squeezing parameter can then be figured out with high precision. We have built an OPO system experimentally working far below the threshold around the waveband of cesium $D2$ line (852 nm) and measured all the related parameters. Weak squeezing of about (-0.066±0.001) dB is eventually determined. The experimental results can be well explained by the theory and this method is independent on the detection efficiency.

## 2. Theoretical analysis

Optical parametric oscillator (OPO) is a workhorse for quantum source generation and it has been investigated theoretically and experimentally[35-36]. When OPO operates near threshold, optimum squeezed vacuum state is generated and it shows even photon number distribution. On the other hand when the OPO works far below the threshold, two-photon state can be prepared. The down-conversion rate of the OPO can be calculated as[37]

$$R = \varepsilon^2 \tau_F F^2 / \pi F_0, \tag{1}$$

where $\varepsilon$ is the single-pass parametric amplitude gain, which is proportional to the pump amplitude and the nonlinear coefficient. $\tau_F = l/c$ is the round-trip time of the OPO cavity with $l$ the cavity length and $c$ the speed of light. $F = 2\pi / \tau_F(\gamma_1 + \gamma_2)$ and $F_0 = 2\pi / \tau_F \gamma_1$ are the finesse of the cavity with and without the loss $\gamma_2$, which $\gamma_1 = T/\tau_F$ with $T$ the output coupler transmission of OPO and $\gamma_2 = L/\tau_F$ with $L$ the extra-losses of OPO. It indicates that the down-conversion rate is linearly proportional to the pump power since $\varepsilon$ is proportional to the pump amplitude. The down-conversion rate can also be expressed as $R = kP$, with $P$ the pump power of OPO and $k$ the linear factor of the pump power. The down-conversion rate can be determined by measuring the number of photons generated from the OPO. The light output from the OPO operates far below the threshold shows a photon super-bunching effect. People generally use the second-order degree of coherence at zero time delay, $g^{(2)}(0)$, to describe the photon statistical properties of light fields. $g^{(2)}(0)$ is equal to the normalized intensity correlation function at zero time delay, which can be calculated as[38]

$$g^{(2)}(0) = 2 + \frac{(\gamma_1 + \gamma_2)^2}{\varepsilon^2}. \tag{2}$$

According to Eqs. (1), (2) and $R = kP$, we can obtain the relationship between the $g^{(2)}(0)$ and the pump power of OPO in weak pump regime, which is expressed as

$$g^{(2)}(0) = 2 + \frac{(\gamma_1 + \gamma_2)^2}{\frac{4\pi F_0}{\tau_F F^2} kP}. \tag{3}$$

We know that people generally describe the output field of OPO by the squeezed vacuum state when OPO operates below the threshold, the output noise variances of the squeezed quadrature ($V_-$) and anti-squeezed quadrature ($V_+$) are given by[39]

$$V_\pm = e^{\pm 2r} = 1 \pm \eta_{esc} \frac{4\sqrt{P/P_{th}}}{(1 \mp \sqrt{P/P_{th}})^2 + 4\Omega^2}, \tag{4}$$

where $r$ is the squeezing parameter. $P_{th}$ is the oscillation threshold of the OPO. $P_{th} = \frac{(T+L)^2}{4E_{NL}}$, where $E_{NL}$ is the single pass conversion coefficient. $\eta_{esc}$ is the escape efficiency of the cavity, which is defined as $\eta_{esc} = T/(T+L)$, $\Omega = 2\pi f/(\gamma_1 + \gamma_2)$ is the normalized measurement frequency with $f$ being the measurement frequency. Therefore, from Eqs. (3) and (4), we can obtain the relationship between the squeezing parameter and the second-order degree of coherence in the weak pump regime, which is expressed as

$$r = \frac{1}{2}\ln\left[1 + \frac{\frac{4\gamma_1 c}{(\gamma_1+\gamma_2)^2 l}\sqrt{\frac{2\gamma_1 E_{NL}}{k(g^{(2)}(0)-2)}}}{\left[1 - \frac{c}{(\gamma_1+\gamma_2)l}\sqrt{\frac{2\gamma_1 E_{NL}}{k(g^{(2)}(0)-2)}}\right]^2 + \left(\frac{2\pi f}{\gamma_1+\gamma_2}\right)^2}\right] \quad (5)$$

From Eq. (5) we can see that squeezing parameter $r$ is related to the second-order degree of coherence $g^{(2)}(0)$. It can be determined when the experimental parameters $\gamma_1$, $\gamma_2$, $l$, $E_{NL}$ and $k$ are measured. Since in the weak pump regime the output of OPO has strong bunching effect and it can be measured easy with high precision, the squeezing parameter is thus can be figured out.

## 3. Experiment setup and results

A schematic diagram of the experimental setup is shown in Fig. 1. We use an 852 nm tunable CW Ti:Sapphire solid laser produced by Msquare Co., and the laser frequency is locked to the cesium atom *D*2 line by polarization spectroscopy method. The crystal used in the second harmonic generator (SHG) and OPO is the type-I PPKTP crystal produced by Raiol Co., and the sizes are 1 mm ×2 mm × 20 mm in SHG and 1 mm ×2 mm × 10 mm in OPO respectively. The crystal is placed in the center of the waist of the cavities and the temperature is precisely controlled to the optimum phase matching temperature. The two cavities are composed of two planar mirrors and two concave mirrors. The curvature radiuses of the concave mirrors are 50 mm in OPO, and the distances between the concave mirrors is 59 mm in OPO, and the waist of OPO is 22.7 μm. The transmission of the coupler mirror is 11%, and a piezoelectric (PZT) element is bonded to one concave mirror, which is used to lock the cavity length. The single pass conversion coefficient ($E_{NL}$) of the crystal is 2% W$^{-1}$. The generated 426 nm blue light, which single pass through the crystal, is used to pump the OPO, and the matching efficiency is 85%. We have used a mode-cleaner (MC), a three-mirror ring cavity, to reduce the intensity noise of the laser. The fineness of the MC cavity is 2000 and the linewidth is 1.4 MHz. The 852 nm light filtered by the MC is used for the local light and the probe light of the OPO. In order to lock the OPO without importing the locking beam into the single photon counting module (SPCM) in the Hanbury-Brown and Twiss (HBT) detection system, we have used an 894 nm laser as the locking beam. This locking beam is locked to the cesium atom $D$1 line by polarization spectroscopy method. In the later we use an 852 nm narrow bandpass interference filter to remove the 894 nm laser light. We use an avalanche photodiode (APD, C30659-900-R5B) to record the faint light signal. By changing the modulation frequency, the sideband frequency of the 894nm locking beam and the 852 nm probing beam are both resonated to the cavity simultaneously and locked by Pound-Drever-Hall (PDH) technology[40].

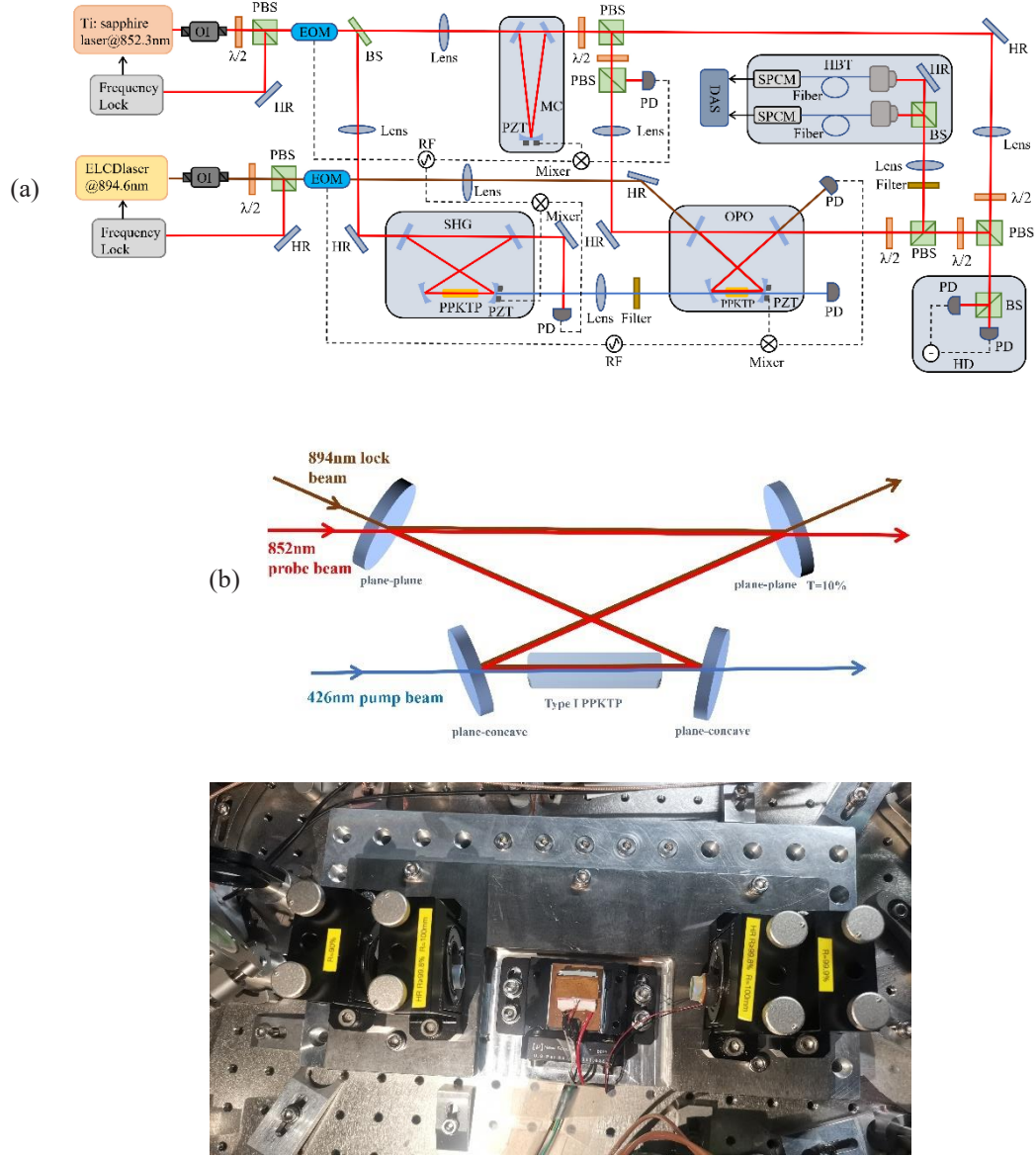

**Fig. 1.** (a) Schematic diagram of experimental setup. OI: optical isolator; $\lambda/2$: half-wave plate; PBS: polarization beam splitter; BS: beam splitter; EOM: electro-optical modulator; HR: high reflectivity mirror; DAS: data acquisition system. The dash lines indicate the control electronics for the cavity length

locking process. PD: photodetector; RF: radio-frequency source. (b) The configuration of OPO and a picture of real system.

The squeezed light generated from the OPO is divided into two parts: one goes to the HD scheme the other to the HBT scheme. First, we have studied the squeezed light generated below the OPO threshold. Different from the output of the OPO, the detected noise variances of the squeezed quadrature ($V_-^{det}$) and anti-squeezed ($V_+^{det}$) quadrature are given as follows [39]

$$V_\pm^{det} = 1 \pm \eta_{det}\eta_{esc} \frac{4\sqrt{P/P_{th}}}{(1 \mp \sqrt{P/P_{th}})^2 + 4\Omega^2}, \qquad (6)$$

where $\eta_{det}$ is the total detection efficiency of the HD: $\eta_{det} = \eta_{tr} \times \eta_{vis}^2 \times \eta_{qu}$, with $\eta_{tr} = 0.95$ the propagation efficiency, $\eta_{esc} = 0.7$ the escape efficiency of the cavity, $\eta_{vis}^2 = 0.97^2$ the interference efficiency, and $\eta_{qu} = 0.99$ the quantum efficiency of the photodiodes. In the experiment, the detected normalized noise power as a function of pump power is shown in Fig. 2. The black squares indicate the experimental anti-squeezed noise data, and the blue squares indicate the experimental squeezed noise data. The red curve represents the theoretical results from the Eq. (6). The quantum noise limit reference is recorded using local power of 2 mW and a blocked probe beam port. The oscillation threshold of the OPO is 165.3 mW. We can see that when the pump power is getting weak, the corresponding squeezing degree is hard to be determined by this traditional HD method.

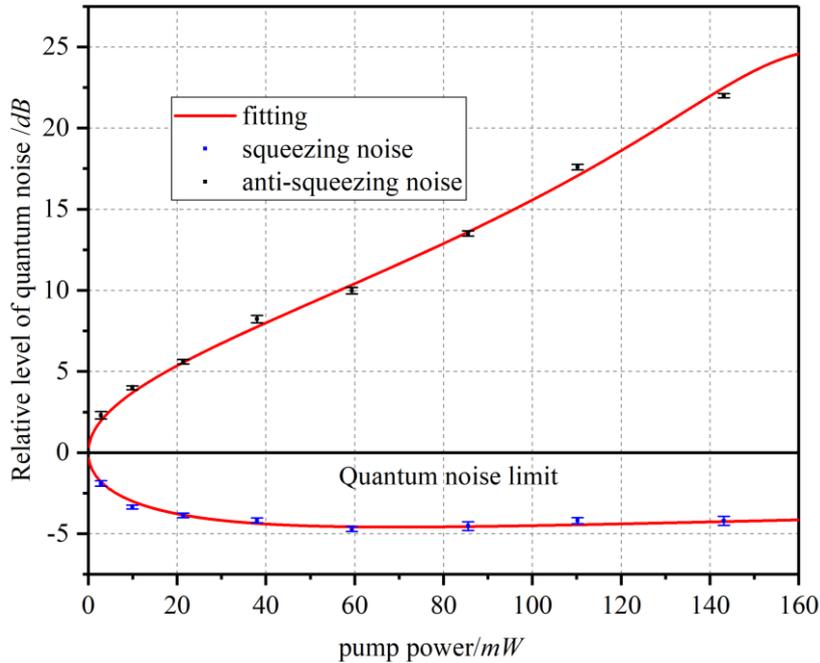

**Fig. 2.** Squeezing and anti-squeezing versus the pump powers. The solid curve represents the theoretical results, and the squares denote experimental results. Analysis frequency is 800 kHz, RBW = 100 kHz and VBW = 30 kHz.

We investigate the photon statistics of the OPO output in the case of low pump regime and the parameters of $\gamma_1$, $\gamma_2$ and $l$ are all determined from the comb-like structure of the second order degree of coherence. The output of the OPO passes through an 852 nm narrow bandpass interference filter and is divided into two paths by a 50/50 beam splitter (BS). The outputs are coupled into two SPCMs through optical fiber. The maximum count rate of the SPCM is 25M counts/s, and the quantum efficiency of the SPCM at 852 nm is about 50%. The dark count is less than 1k counts/s. The outputs of the SPCMs enter into a data acquisition system (QuTag). The second-order degree of coherence in this case is expressed as[38]

$$g^{(2)}(\tau) = N_1 \left[ N_2 + e^{-\Omega_c|\tau-\tau_0|} \sum_n \left(1 + \frac{2|\tau - n\tau_F - \tau_0|\ln 2}{\tau_R}\right) \times \exp\left(-\frac{2|\tau - n\tau_F - \tau_0|\ln 2}{\tau_R}\right) \right]. \quad (7)$$

Here $N_1$ and $N_2$ are constants which are proportional to the pump power. $\tau_0$ denotes an electronic delay, $n$ the number of nondegenerate modes, $\Omega_C/2\pi = (\gamma_1 + \gamma_2)/2\pi$ the linewidth of the OPO, and $\tau_R$ resolution time of detection system. It should be noted that second-order degree of coherence presents the comb-like structure with a peak interval $\tau_F = 2\pi/\Omega_F$. The measured experimental result of $g^{(2)}(\tau)$ is shown in Fig.3. The OPO pump light power is 30 μW. The resolution time of the DAS is 35 ps, and the bin count is 4000. The black squares indicate the experimental data, and the blue curve represents the theoretical fitting result based on the Eq. (7). From the theoretical fitting we get the following parameters: $\gamma_1$ = 82.1 MHz, $\gamma_2$ = 6.9 MHz, and $l$ = 405 mm. It should be noted that the important parameters of the OPO, such as the total losses, corresponding to the linewidth, are determined by this measurement. We get the total losses 11.5% of the OPO and the cavity linewidth 14.16 MHz with the cavity length 405mm.

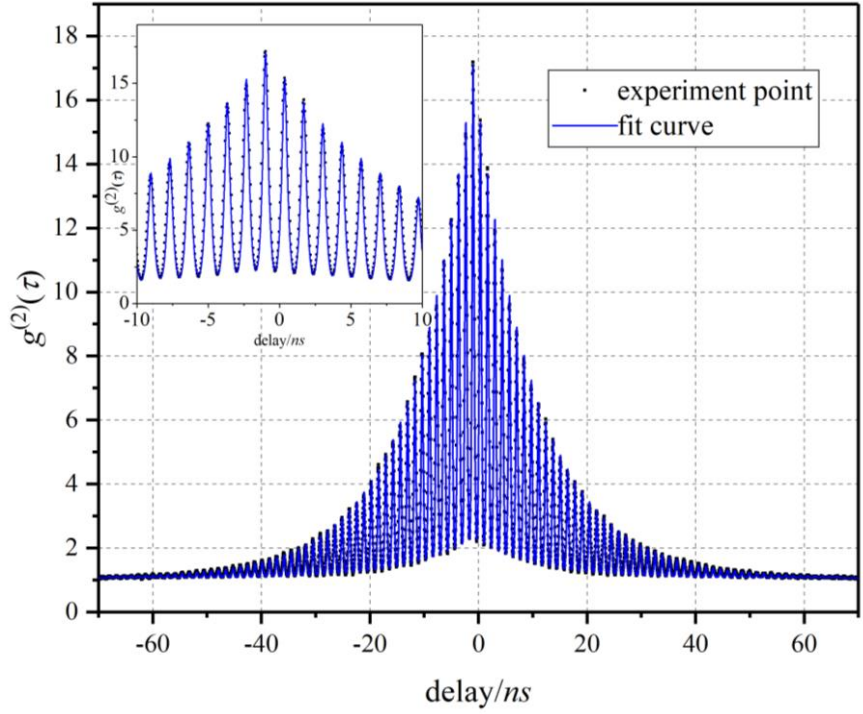

**Fig. 3.** Experimental result of second-order degree of coherence. The solid blue line represents the theoretical fitting and black squares denote experimental results. The inset shows the results of delay time from -10 ns to 10 ns. The pump power is 30 μW. The fitting parameters are as follows: $N_1 = 16$, $N_2 = 0.064$, $\Omega_C/2\pi = 14.16$ MHz, $\tau_0 = -0.98$ ns, $\tau_R = 185$ ps, and $\tau_F = 1.34$ ns.

For determining the $k$ in Eq. (5), we have measured the photon generation rate as the pump power varies. Fig. 4 shows the measured down-conversion rate $R_{meas}$ varies with the pump power. The solid line represents the theoretical fitting and the blue dots denote the experimental results. The results show that $R_{meas}$ increases linearly according to the pump power. The generated rate of down-conversion by OPO, $R = R_{meas}/\eta$, can be calculated by $R_{meas}$ considering the detection efficiency $\eta$, which is given by $\eta = tfd$. In our experimental system, the transmittance from OPO to fiber is $t = 0.85$, the fiber coupling efficiency is $f = 0.95$, and the SPCM efficiency is $d = 0.5$. The background noise is also deducted. From the linear fitting we get $k = 104.5$ MHz/mW.

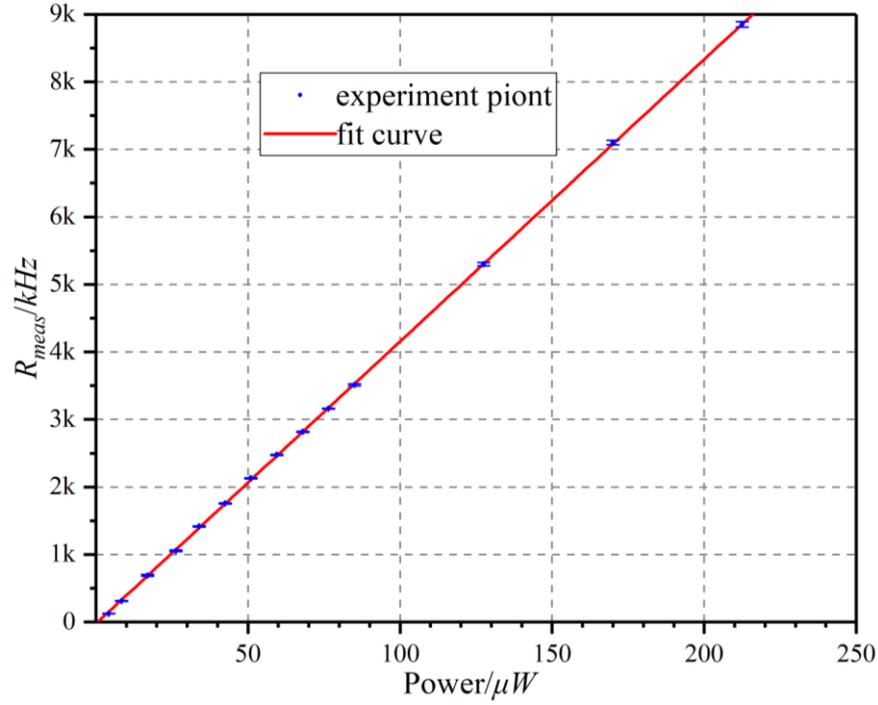

**Fig. 4.** Measured down-conversion rate $R_{meas}$ versus pump powers with detection efficiency $\eta = 0.4$. The solid line represents the theoretical fitting and the blue dots denote the experimental results. The error bar is the statistical error from 5 measurements.

As expected, the $g^{(2)}(0)$ is inversely proportional to the pump power according to Eq. (3). We have measured $g^{(2)}(0)$ versus pump powers for certain detection efficiency $\eta=0.4$. The photon statistics of the OPO working far below threshold shows strong bunching effect and the $g^{(2)}(0)$ can be measured accurately by HBT scheme. The result is shown in Fig. 5. The theoretical fitting agrees well with the experimental data. We also measure the $g^{(2)}(0)$ with the pump power under different efficiencies. The result is shown in Fig. 6. The results indicate that the measured $g^{(2)}(0)$ is almost unchanged at fixed pump power when the detection efficiency varies. Even if the total detection efficiency is dropped to $0.2\eta$, the system still works for $g^{(2)}(0)$ measurement.

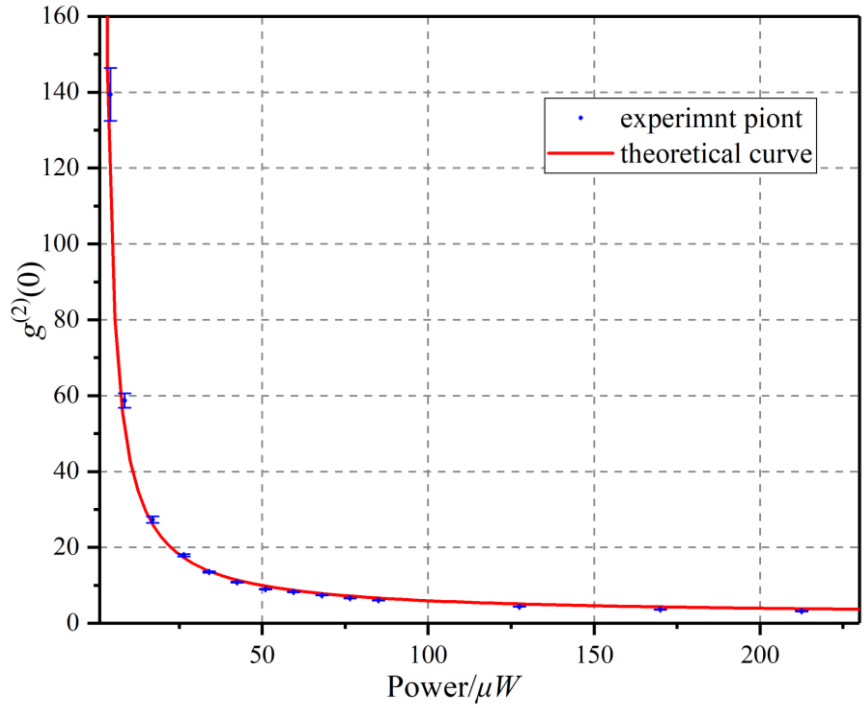

**Fig. 5.** Measured $g^{(2)}(0)$ versus pump powers with detection efficiency $\eta = 0.4$. The red solid line represents the theoretical fitting and the blue dots denote the experimental results. The error bar is the statistical error from 5 measurements.

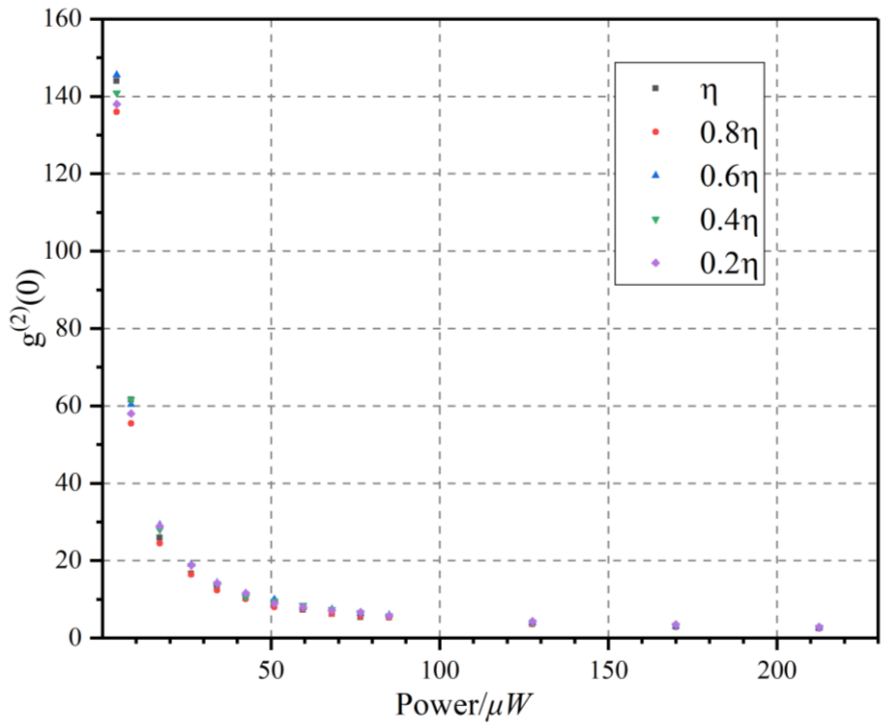

**Fig. 6.** Measured $g^{(2)}(0)$ versus pump powers under different total detection efficiencies with the detection efficiency $\eta = 0.4$. The colored dots denote the experimental results.

Based on the above measurements, we eventually obtain the results of squeezing parameter $r$ and the $g^{(2)}(0)$ in the case of far below the OPO threshold, and the results are shown in Fig. 7 (left side) when the pump power is from 5 μW to 200 μW. The solid line represents the theoretical fittings according to Eq. (5) and the points are the experimental results. All the parameters are determined according to the experimental data. The theoretical analysis is consistent well with the experiment. Since the photon statistics is very sensitive to the squeezing which relies on the pump rate, very small change of the pump rate can result in a great amount of change of the $g^{(2)}(0)$. We can then infer to the tiny change of the squeezing by $g^{(2)}(0)$ measurement. The right side of Fig.7 shows the squeezing in dB scale. Weak squeezing of (-0.066±0.001) dB is eventually determined when the pump is 5μW. In the case of 200 μW of the pump power the squeezing is (-0.54±0.01) dB. Comparing to the general HD measurement, there are two advantages. Firstly, the method provides an alternative way to determine the weak squeezing that goes beyond to the measured limitation by usual HD detection. Secondly, such approach to determine the squeezing parameter $r$ is independent of the detection efficiency.

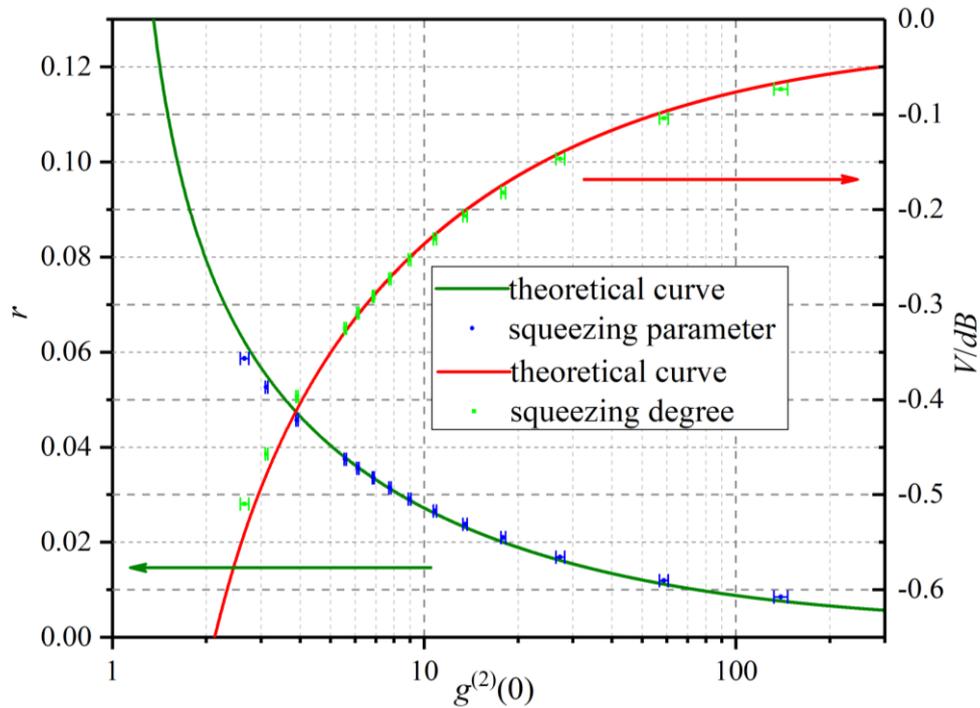

**Fig. 7.** Squeezing parameter $r$ versus the second-order degree of coherence with zero time delay $g^{(2)}(0)$. The left ordinate in the figure is the squeezing parameter, and the right coordinate is the squeezing degree in dB scale. The solid line represents the theoretical fittings and the dots denote the experimental results with error bar from 5 measurements. The corresponding fitting parameters: $\gamma_1$ = 82.1 MHz, $\gamma_2$ = 6.9 MHz, $E_{NL}$ = 2% W$^{-1}$, $l$ = 405 mm, $k$ = 104.5 MHz/mW and $f$ = 800 KHz.

4. **Conclusion**

We have presented a feasible method to determine the weak squeezing parameter of the OPO based on the measurement of photon statistics. The relationship between the second-order degree of coherence and the squeezing parameter is established. Experiment is done by setting up an OPO system working at the wavelength around the 852nm. By intentionally controlling the pump rates and the total detection efficiencies, we have measured the second-order degree of coherence from OPO on various experimental situations, and proved that weak squeezing, along with strong bunching of photon statistics, can be determined. This approach appears advantages either for extremely weak non-classicality verification or the detection-efficiency-free measurement of quantum light sources. The measure based on photon statistics may be extended to even higher order correlation to reveal other quantum features of light field[41-42].

**Acknowledgement**

This work was supported by the National Natural Science Foundation of China (Grant Nos. 11634008), the National Key Research and Development Program of China (Grant No. 2017YFA0304502).